\documentclass[12pt]{book}
\usepackage{epsf}
\usepackage{wrapfig}
\usepackage{graphicx}
\usepackage{amssymb}
\usepackage{amsmath}
\usepackage{color}
\usepackage[utf8]{inputenc}
\usepackage[english]{babel}

\begin{document}

\title*\begin{center}
{\textbf{{\LARGE \ On  the $ \rho $ and $ \sigma_{tot} $ measurement by
the TOTEM Collaboration: in the wake of recent discoveries.}}}
\end{center}

\begin{center}
V. V. Ezhela\footnote{Vladimir.Ezhela@ihep.ru},
V. A. Petrov\footnote{Vladimir.Petrov@ihep.ru} and
N. P. Tkachenko\footnote{Nikolai.Tkachenko@ihep.ru}
\end{center}
\begin{center}
 A.A. Logunov Institute for High Energy Physics,\vspace{-4.0mm}
\begin{center}
NRC "Kurchatov Institute", Protvino, RF
\end{center}
\end{center}

\begin{center}
Abstract\vspace{-3.0mm}
\end{center}
{\it We show that extraction of the quantities}
$\rho(s)=\mbox{Re}[T_{N}(s,0)]/\mbox{Im}[T_{N}(s,0)]$ {\it and}
$\sigma_{\mbox{tot}}(s)$ {\it from the data
on the $pp$ differential cross-section at $\sqrt{s} = 13$} TeV {\it obtained
by the} TOTEM {\it Collaboration gives results essentially different
from those presented in  publication} \cite{Ant} {\it if to use a modified
formula for Coulomb-nuclear interference. The physical interpretation of these data changes accordingly.}\vspace{-4.0mm}

\section*{The Problem.\vspace{-3.0mm}} 

  The experimental data of the TOTEM group on differential,
$(d\sigma/dt)(s,t)$,  and total, $\sigma_{\mbox{tot}}(s)$,
cross-sections  in $pp$-scattering,
as well as the parameter $\rho (s)\equiv \mbox{Re}[T_{N}(s,0)]/\mbox{Im}[T_{N}(s,0)]$
at the LHC energy  $ \sqrt{s}=13$ TeV, were published in \cite{Ant}.
 
 An unexpectedly small (compared with the predictions of most models) value
of the parameter ($\rho= 0.1$ or even less) was interpreted - based on a
comparison with one specific model \cite{MaNi}  - as the discovery of
the "maximal Odderon" supposedly in the form of a "3-gluon bound state" or,
"at least", as
{\it slowing down of the growth of total cross sections in the} LHC
{\it energy region}. Also noteworthy is quite a categorical statement that
{\it "the unprecedented precision of the $ \rho $ measurement, combined with
the} TOTEM {\it total cross-section measurements in an energy range larger
than  10} TeV ({\it from} 2.76 {\it to} 13 TeV) {\it has implied the exclusion
of all the models classified and published by} COMPETE"\footnote{This refers
to a theoretical collaboration that published in \cite{COMP} a general
parameterization of the total cross sections and parameters $\rho$ for
all available initial channels.} (italics are ours).

  Let us remind that the "maximal Odderon" doctrine was launched in
 Ref.\cite{Nic}  where it was assumed that the upper bound for high-energy
 behaviour of the  crossing-odd forward scattering amplitude is  functionally
 saturated and  leads eventually to a maximal violation of the difference
 form of the Pomeranchuk  theorem which predicted, in particular, that\vspace{-3.0mm}
$$\sigma_{\mbox{tot}}^{pp} -\sigma_{\mbox{tot}}^{\bar{p}p}
\rightarrow 0\vspace{-3.0mm}$$
at high energies while the "maximal Odderon" option implies instead
that at
$s\rightarrow\infty$\vspace{-3.0mm}
$$\sigma_{\mbox{tot}}^{pp} -\sigma_{\mbox{tot}}^{\bar{p}p} \sim +\ln s.
\vspace{-3.0mm}$$
It is not surprising that the results \cite{Ant} being related , as we see,
to the most important conceptual problems of the theory of strong interactions
at high energies naturally aroused considerable interest and a number of
discussions, accompanied by several dozen publications\footnote{These results
made a fuss in the media \cite{CEc} including the tabloid press \cite{News}.}.
 
In this regard, we believe that a thorough analysis of the results obtained in\cite{Ant} (and the physical conclusions made on their basis), undertaken in the present work (including their re-analysis on the base of different theoretical means), seems quite appropriate and at least not useless, especially on the eve of the upcoming measurements at the LHC energy of 14 TeV.\vspace{-6.0mm}

 \section*{Dealing with the TOTEM data.\vspace{-3.0mm}} 
 
\begin{figure}[h]
$$\includegraphics[width=170mm]{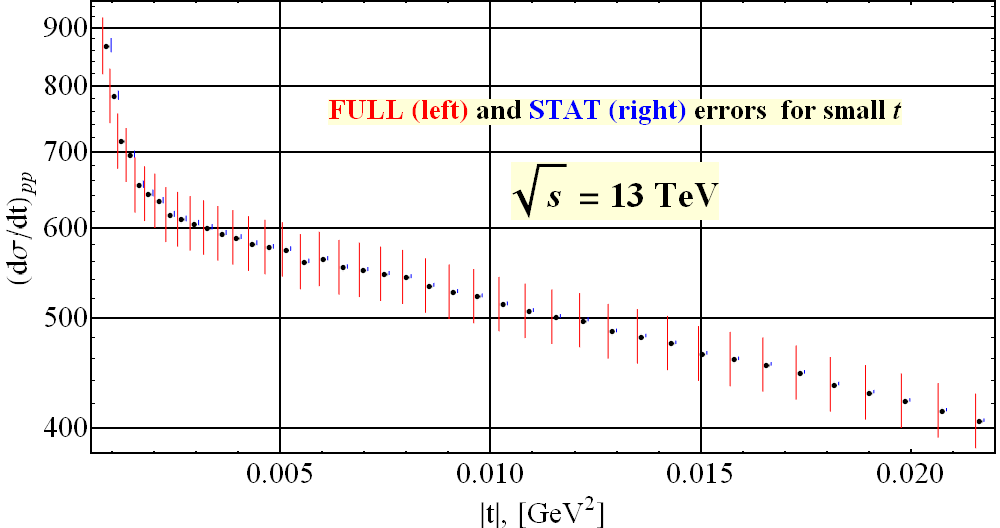}\vspace{-6.0mm}$$
\caption{Systematic and statistical errors of $ (d\sigma/dt)(s,t)$
in the TOTEM experiment at low momentum transfers (double logarithmic scale).}
\label{fig-1}       
\end{figure}

 When acquainting with experimental data on the differential cross sections of $ pp $ scattering at 13~TeV obtained by the TOTEM
collaboration \cite{Ant} (Table 3 in that article), the fact is noteworthy that in many cases the systematic errors in determining the value of
$(d\sigma/dt)(s,t)$  exceed the statistical errors an order of magnitude
or more. This fact is especially pronounced for small values of
$|t|$ as shown in Fig.\ref{fig-1}.

Approximately half of the experimental data (low $|t|$) are indicated with
a detailed indication of the sources of errors(see Table 3 in \cite{Ant}),
but the remaining (larger $|t|$) data (see Table 5 in \cite{TOT'}) contain
only statistical and systematic errors as a whole. Therefore, it is impossible
to build a correlation matrix between the measurement results. There is nothing
else left but to assume, for a while, the total measurement error to be
$\sqrt{\mbox{stat}^{2}+ \mbox{syst}^{2}}$.
The TOTEM collaboration obtained values of $(d\sigma/dt)(s,t)$  down to
very small $|t|\geqslant 0.000879~\mbox{GeV}^{2}$ which allows one to
estimate the value of the parameter
$\rho (s)= \mbox{Re}[T_{N}(s,0)]/\mbox{Im}[T_{N}(s,0)]$ when  fitting experimental
data only for very small values of the transferred momentum.
To conduct further quantitative estimates, it is necessary to precise  what
is meant by "small values of the transferred momentum" ($|t|$).
Specifically, all experimental values are being used at
$|t|\leqslant t_{0}$ while discarding all experimental values at
$|t| >t_{0}$. According to the set of experimental data selected in
this way, fitting was carried out using the model for the strong interaction
amplitude given in \cite{Ant} (the "TOTEM model") and making sequential monitoring
of the dependence on $t_{0}$.\vspace{-4.0mm}
\newpage
 
\section*{Notations, normalization and theoretical
requisite}\vspace{-3.0mm} 
 
 In general case, the differential cross section is expressed in terms of
 the scattering amplitude $ T_{N + C} (s, t) $ ( in this paper we consider
 only $pp$ scattering) which depends on the c.m.s. energy $\sqrt{s}$
and the transferred momentum squared $t$:\vspace{-2.0mm}
{\boldmath
$$
\frac{d\sigma}{dt}(s,t)=
\frac{g |T_{N + C} (s, t)|^{2}}{16\pi s(s-4m_{p}^{2})}.
\vspace{-2.0mm}
$$
}
\noindent Index $N+C$ means account both of strong (N, "nuclear") and
Coulombic (C) forces while\linebreak
$g= (\hbar c)^{2}~[\mbox{GeV}^2 \mbox{mb}]$ is the usual conversion factor
(the amplitude is  dimensionless).
The total cross section $\sigma_{\mbox{tot}}(s)$ (till the end of the paper we will
omit the index "\textit{tot}") and
$\rho(s,t)= \mbox{cot}^{-1} [\mbox{Arg}T_{N}(s, t)]$ are expressed in terms
of the scattering amplitude in the standard way:
\vspace{-1.0mm}
{\boldmath $$\sigma_{\mbox{tot}}(s)= g\cdot \mbox{Im}[T_{N}(s, 0)]/\sqrt{s(s-4m_{p}^{2})},
~~~
\rho(s,t)= \mbox{Re}[T_{N}(s,t)]/\mbox{Im}[T_{N}(s,t)].$$}\vspace{-8.0mm}

In turn, the scattering amplitude itself is expressed in terms of the sum of the nuclear scattering amplitude $T_{N}(s,t)$, the pure Coulomb component
(taking into account the proton form factor\linebreak
$F(t)=1/(1-t/(0.71~\mbox{GeV}^{2})^{2}$) and two additional terms
$L_{1}(s,t)$ and  $L_{2}(s,t)$ as was obtained in \cite{Ca} and then invariably
used in the TOTEM publications:
\begin{equation}
\boldmath{
T_{N+C}(s,t)= T_{N}(s,t) + \frac{8\pi \alpha s F^{2}(t)}{t}+
L_{1}(s,t)+ L_{2}(s,t).} \label{eq1}
\end{equation}
The last two terms reflect the Coulomb-nuclear interference at the amplitude level and are expressed by the formulas:\vspace{-2.0mm}
$$
\begin{array}{l}
L_{1} (s,t)= \frac{i\alpha }{\pi }
\int_{-\infty}^{0} dt' I(t,t')
\left[ T_{N}(s,t)- T_{N}(s,t') \right] ,\\[2mm]
L_{2}(s,t)= - i\alpha T_{N}(s,t) \int_{-\infty}^{0}dt'
\ln\left(\frac{t^{'}}{t}\right)
\frac {dF^{2}(t^{'})}{dt^{'}}
\end{array}
$$
where
$$I(t,t')= \int^{(\sqrt{-t}+\sqrt{-t'})^{2}} _{(\sqrt{-t}-\sqrt{-t'})^{2}}
\frac{F^{2}(x)}{\sqrt{((\sqrt{-t}+\sqrt{-t'})^{2}-x  )(x-(\sqrt{-t}-\sqrt{-t'})^{2})}}dx.$$

Note that the above relations are of a general nature and do not imply any
model expressions for the amplitude $ T_{N}(s,t) $ which can be selected
in various ways already during the data processing.
It must be pointed out that in expression (\ref{eq1}), as shown in [3], the
term $L_{2}(s,t)$ is redundant. However, since the TOTEM group still uses
an expression containing this term, we present fitting options with both
the single term $L_{1}(s,t)$ and with addition of $L_{2}(s,t)$. The latter
is necessary in order to trace the source of the numerical value for
$\rho$ obtained in \cite{Ant}.
It is interesting that if interference at the amplitude level is not taken
into account at all, no terms $ L_{1,2}(s,t) $ (which is formally, of course,
incorrect), then the result of the fit — in any case for those "nuclear"
amplitudes that are used below — is close to "option 1" (only $ L_{1}(s,t) $). In particular, this suggests that with the "nuclear" amplitude according to the "TOTEM model" (see the next Section), the interference term $  L_{1}(s,t) $ is numerically small in the region of low transfers and at
$\sqrt{s} = 13$ TeV, which undoubtedly affects the numerical estimate of
parameter $\rho$. More on this in the next section\footnote{In the general
case, for fitting experimental data on a complete set of experimental values
of $t$\newline
($\mid t \mid \in [ 0.00088 \div \cong 5]~\mbox{GeV}^{2}$), we use
the tested model described in \cite{Av}.}.\vspace{-7.0mm}

\section*{"TOTEM model" and the experimental data at small $t$.\vspace{-3.0mm}}
 
For small transfers, formula (\ref{eq1}) is still valid with the refinement that
the expression for the nuclear amplitude at $ t\rightarrow 0 $ can be significantly
simplified, and we adopt the expression for $ T_{N}(s,t)$ in the form used
by the TOTEM collaboration ("TOTEM model") for processing experimental data
at small $t$, namely\vspace{-5.0mm}
\begin{equation}
\boldmath{
T_{N}(s,t)= 4s(i+ \rho)\sqrt{\frac{\pi}{1+\rho^{2}}\frac{1}{g} \frac{d\sigma}{dt}(s,0)}
~e^{( b_{1}|t|+ b_{2}|t|^{2} + b_{3}|t|^{3})/2 }}\label{eq2} \vspace{-1.0mm}
\end{equation}
where the parameters $\rho$, $\frac{d\sigma}{dt}(s,0)$, $b_{1,2,3}$
are determined from fitting the experimental data.
In \cite{Ant}, experimental data on differential cross sections are presented,
including small values of the transferred momentum (down to
$10^{-3}~ \mbox{GeV}^{2} $). On this basis, parameters, such as
$\rho$, are determined when fitting the experimental data at the lowest possible
values of $ t $. In fact, the experimental TOTEM data are presented in two
versions (Table 3 in \cite{Ant}):\vspace{-3.0mm}
\begin{enumerate}
\item Differential cross-sections with a systematic and experimental error
(5 and 6 columns in Table 3 in \cite{Ant}). It can be seen that systematic
errors in most cases significantly exceed statistical errors - sometimes
by more than an order of magnitude.\vspace{-3.0mm}
\item Differential sections with a weighted set of error sources of these
values (7-11 columns in Table~3 in \cite{Ant}). In this case, it is also
easy to notice that the weighting coefficients in the first column sharply
exceed the coefficients in the remaining columns.\vspace{-3.0mm}
\end{enumerate}

Due to the described properties of the experimental data, we have analysed
the experimental data in the following four options:\vspace{-2.0mm}
\begin{enumerate}
\item $ d\sigma/dt $ with the systematic and experimental errors:\vspace{-3.0mm}
\begin{itemize}
 \item  with the statistical error only;\vspace{-2.0mm}
 \item with the total error.\vspace{-2.0mm}
\end{itemize}
\item $d\sigma/dt$  with a weighted set of error sources:\vspace{-3.0mm}
\begin{itemize}
 \item excluding the source of errors in normalization (8-11 columns in
 Table~3 in \cite{Ant});\vspace{-2.0mm}
 \item taking into account all sources of errors (columns 7-11 in Table~3
  in \cite{Ant}).
\end{itemize}
\end{enumerate}
However, before presenting the results of these four options, it is necessary
to first consider the values of the interference components
$L_{1}(s,t)$ and  $L_{2}(s,t)$ in the main formula (\ref{eq1}).\vspace{-4.0mm}

\begin{figure}[h]
$$\includegraphics[width=140mm]{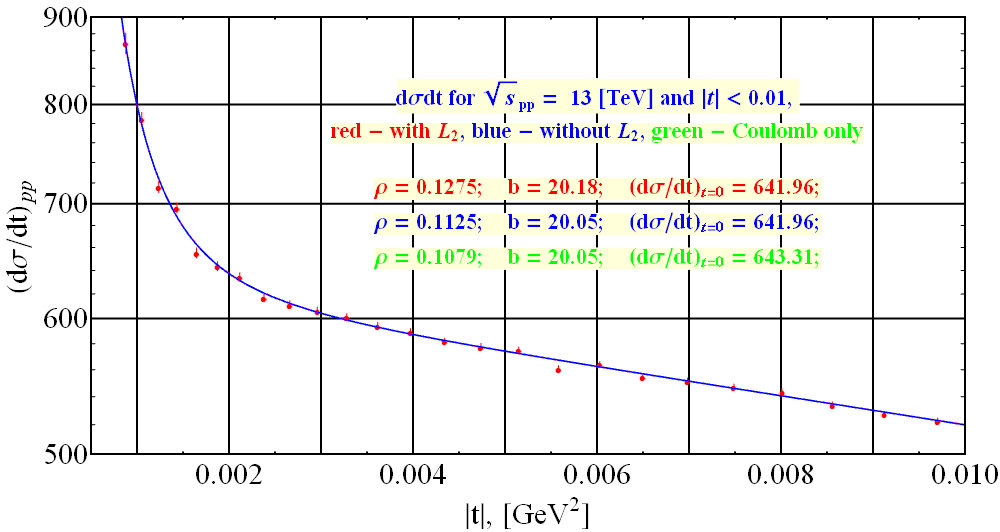}\vspace{-3.0mm}$$
\caption{Experimental data on $d\sigma /dt$ in $pp$ scattering at
$\sqrt{s}= $13 TeV, $|t| < 0.01~\mbox{GeV}^{2}$  and parameter values
for three fitting variants. All curves are graphically indistinguishable.
The graph shows only statistical errors (for systematic see the previous
figure)}
\label{fig-2}       
\end{figure}
\newpage
\section*{Evaluation of various forms of CNI with the same  TOTEM nuclear model.}\vspace{-2.0mm}

We begin by examining the role played by the terms $L_{1}(s,t)$ and
$L_{2}(s,t) $ in expression (\ref{eq1}) for the full amplitude at small $t$.
As mentioned above, all the data will be omitted for $ \mid t\mid $ above
some value $ t_{0} $, and the remaining experimental points are fitted from
the side of small values of $ \mid t \mid $ for various values of
$t_{0}$.

The issue of errors of the fit parameters is left for special consideration
below. 

The fit was carried out in the following three options for the full amplitude:
\begin{enumerate}
\item {\boldmath $T_{C+N}(s,t) = T_{N}(s,t)+
\frac{8\pi \alpha s F^{2}(t)}{t} + L_{1}(s,t)$}
 
In accordance with the above, it is in this form that it is necessary to
take the total scattering amplitude, where $ L_{1}(s,t) $ takes into account
the Coulomb-nucleus interference in the lowest order in $ \alpha $.
\item {\boldmath $T_{C+N}(s,t) = T_{N}(s,t) + \frac{8\pi \alpha
s F^{2}(t)}{t} + L_{1}(s,t)+ L_{2}(s,t)$}

Here we intentionally hold the excess term $ L_2 (s, t) $ so that
to trace the genesis of the $ \rho $ value obtained in \cite{Ant}.\vspace{-3.0mm}
\item {\boldmath $T_{C+N}(s,t) = T_{N}(s,t) + \frac{8\pi \alpha s F^{2}(t)}{t}$}

This (formally incomplete) option was used to evaluate the numerical significance
of the term $ L_{1}(s,t) $ in option 1.
\end{enumerate}
As an example, let us first give the graphic result of the fits that correspond
to these three options (Fig.\ref{fig-2}).

The results of the fitting parameters are extremely close to each other\footnote{However,
they cannot be considered definite until their errors are calculated - more
on that below.} , and the behaviour of the curves is visually indistinguishable
for these three options.
 
\begin{figure}
$$\includegraphics[width=140mm]{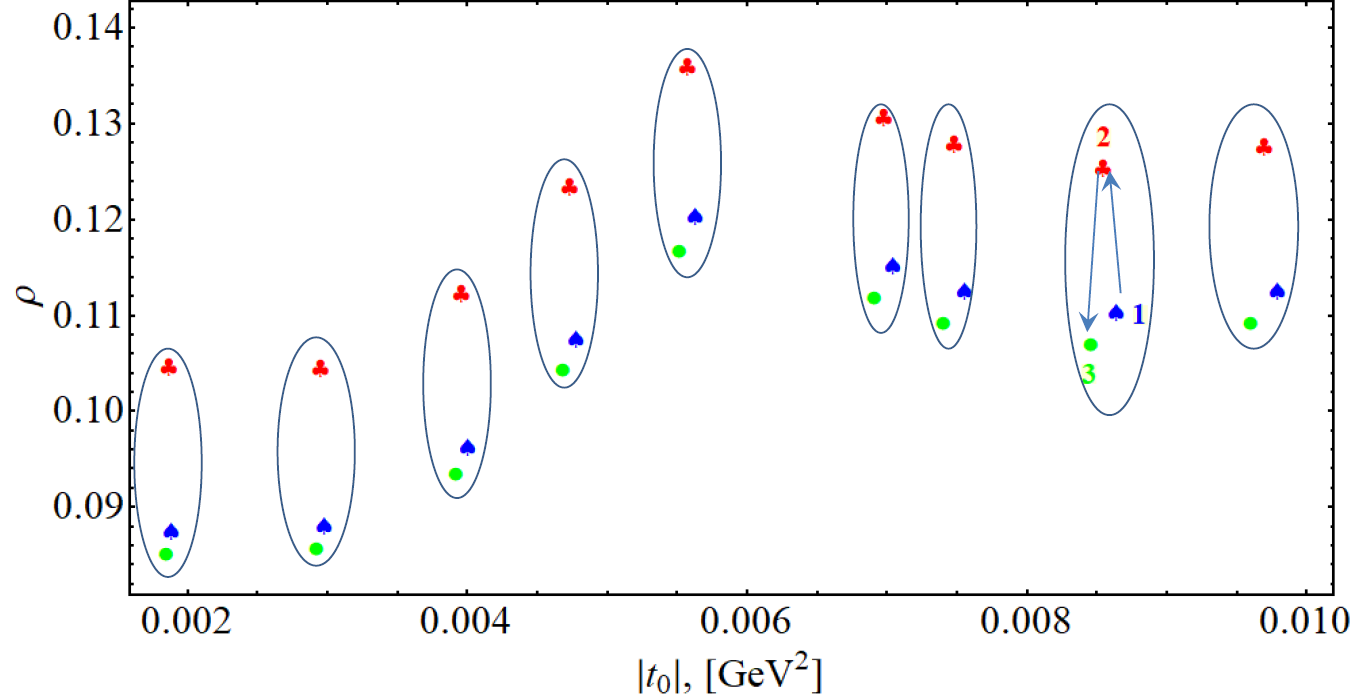}\vspace{-3.0mm}$$
\caption{}
\label{fig-3}       
\end{figure}

The values of the fitting parameters behave identically for these three cases
(depending on the cut-off value $ t_0 $ , and they are presented in
Fig.\ref{fig-3}. 

It can be seen from the figure that if to  go  from the main option 1 (on the graphs
it is indicated by the number 1)to the option with then term
$L_{2}(s,t) $ (option 2 - on the graphs it is indicated by the number 2),
then the parameter values increase significantly. In option 3, i.e. without
using the terms $L_{1}(s,t)$ and $L_{2}(s,t)$, the parameter values practically
return to the original version 1.
\newpage

\section*{Results when considering only statistical and complete errors of experimental data\vspace{-3.0mm}}

\begin{figure}
$$\includegraphics[width=140mm]{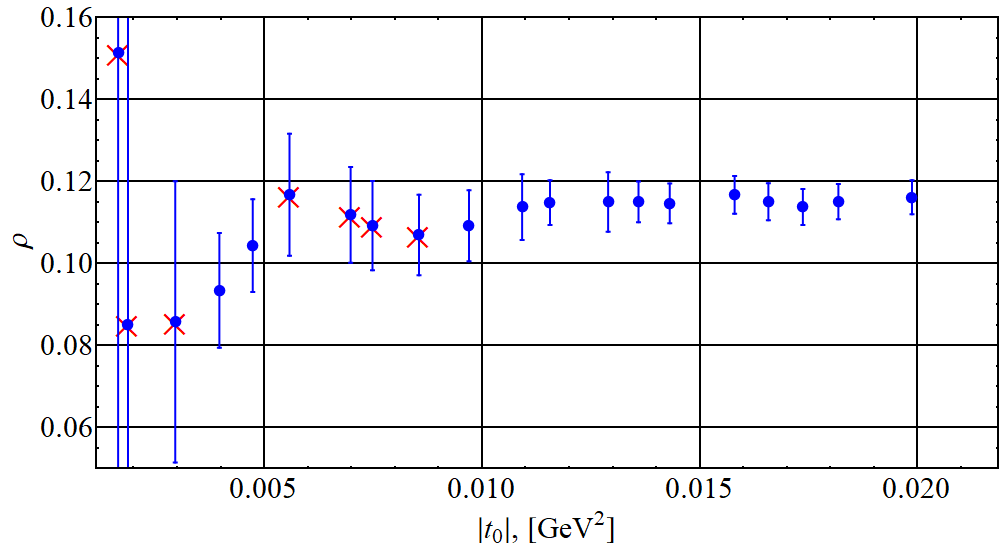}$$
\caption{Values of the $ \rho $-parameter and its error depending on the
cut-off value of the experimental data $ \mid t_{0}\mid $ taking into account
only statistical errors. Crossed out are results for which
$\chi^{2}/\mbox{NoF}>1$.}
\label{fig-4a}      
$$\includegraphics[width=140mm]{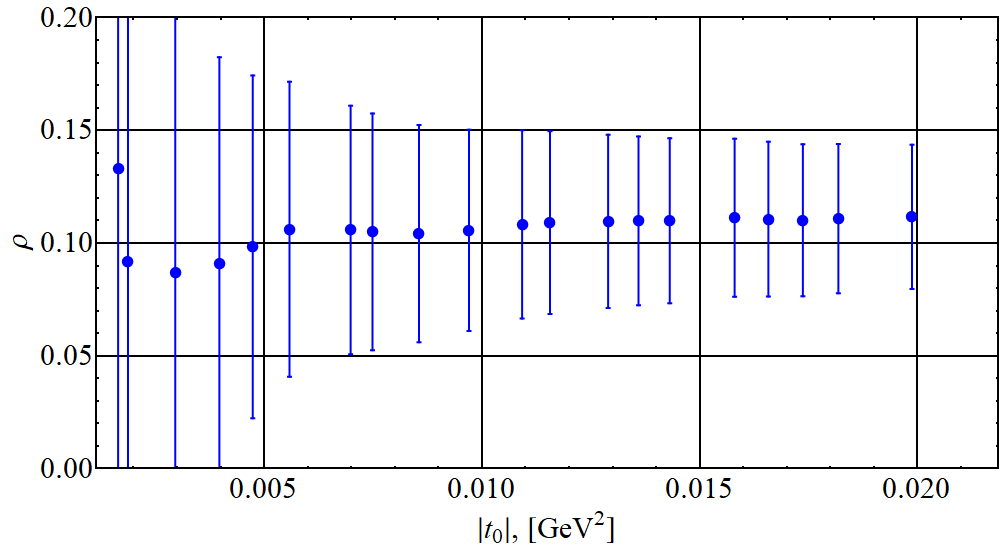}$$
\caption{Values of the $\rho $-parameter and its error depending on the
cut-off value of the experimental data $\mid t_{0} \mid $ with full errors.
However, in all these results, $ \chi^{2}/\mbox{NoF} \ll 1 $, and these
results should be regarded as evaluative.}
\label{fig-4b}       
\end{figure}

Due to the large magnitude of systematic errors for small $|t|$
the value $\chi^{2}/\mbox{NoF}$, when using such experimental data with
a small number of points (as can be seen from the graph in Fig. 1),
will be close to zero even with three fitting parameters (retaining only
$b_1 \equiv b$ from three $b_{i}$), which negates the fit result. In addition,
with a decrease in the number of experimental data, a larger number of fitting
parameters does not allow one to approach small values of $t$ as close
as possible. Three  parameters $b_{i}$ (see Eq. (5)) are redundant, and it
is more correct to use only one parameter $b_1 \equiv b$ (note that in
the TOTEM publication [1] this minimization of  parameters $b_{i}$ was also
used when only statistical errors were taken into account and for
$t_{0}= 0.07~\mbox{GeV}^{2}$).
In the case of higher energies, there is nothing else left but to use
the complete errors of the experimental data
$\sqrt{\mbox{stat}^{2}+ \mbox{syst}^{2}}$.
Thus, from the foregoing, the question arises: how was it possible to obtain
the "experimental value" of the $ \rho $ parameter with an accuracy of about
$10\%$, as stated in the publication [1]?
As noted above, we investigated two options: one with only statistical errors
and another, with complete errors in experimental data.
The behaviour of the parameter $ \rho $ with a decrease in the cut-off value
$t_0$  is shown in Fig. \ref{fig-4a} and Fig. \ref{fig-4b}. It can be seen
that approaching the zero value of $t$ fails closer than to
$|t|\approx{0.004}~\mbox{GeV}^{2}$. In this case, it really turns
out that $\rho\approx 0.09$. However, the total error of this value is
almost $100\%$. The error $\mathcal{O}$$(10\%)$ (as can be seen from the
results given in [1] $\rho =0.09 \pm 0.01$ and $\rho = 0.10 \pm 0.01$)
does not work even if only statistical errors are taken into account: in
fact, they are no less than $20\%$ .

\begin{figure}
$$\includegraphics[width=140mm]{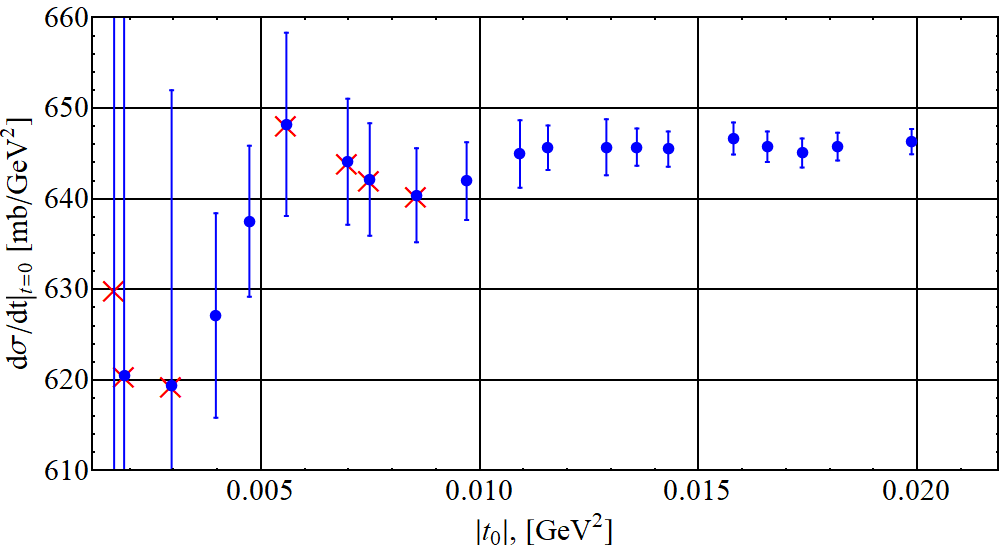}$$
\caption{The values of the parameter ${(d\sigma/dt})_{t=0}$ and
its error, depending on the cut off value $|t_{0}|$ of the
experimental data  taking into account only statistical errors. Crossed
out are the results for which $ \chi^{2}/\mbox{NoF} > 1 $.}
\label{fig-5a}       
$$\includegraphics[width=140mm]{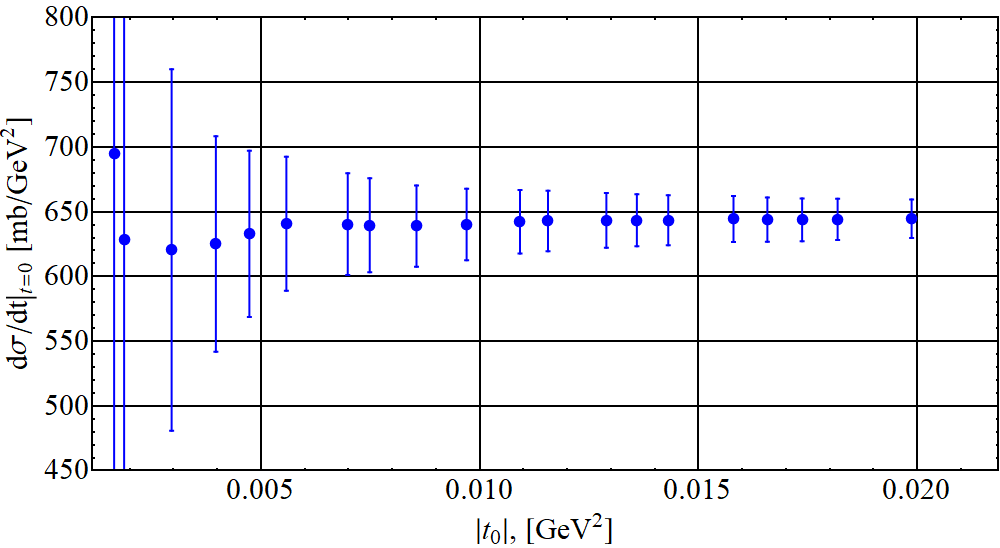}$$
\caption{The values of the parameter $(d\sigma /dt)_{t=0}$  and
its error, depending on the cut-off value $|t_{0}|$ of the
experimental data  with full errors. However, in all these results,
$\chi^{2}/\mbox{NoF} \ll 1$, and these results should be regarded as
evaluative.}
\label{fig-5b}       
\end{figure}

\begin{figure}
$$\includegraphics[width=140mm]{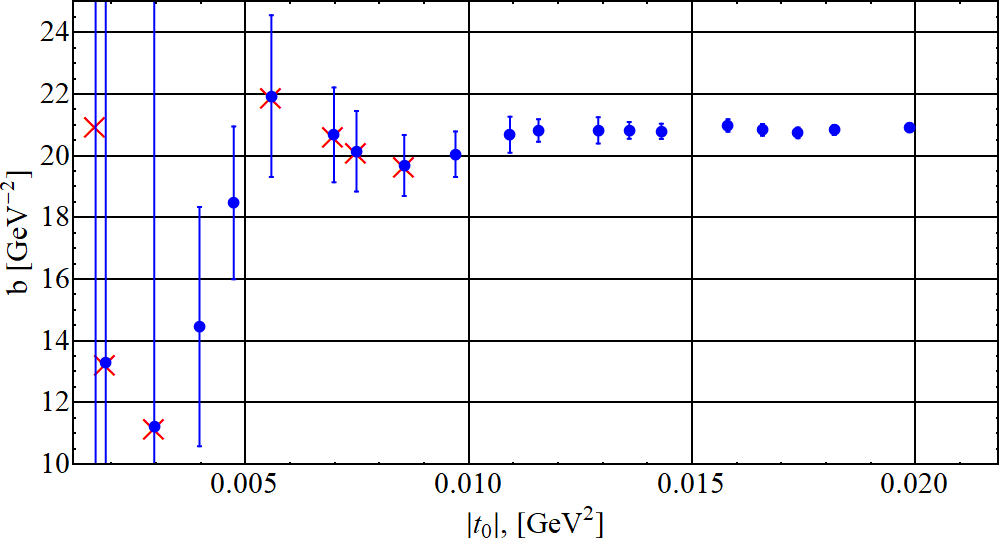}$$
\caption{The values of the parameter $b$ and its error
depending on the cut-off value $|t_{0}|$ of the experimental
data  taking into account only statistical errors. Crossed out are the
results for which $\chi^{2}/\mbox{NoF}> 1$.}
\label{fig-6a}       
$$\includegraphics[width=140mm]{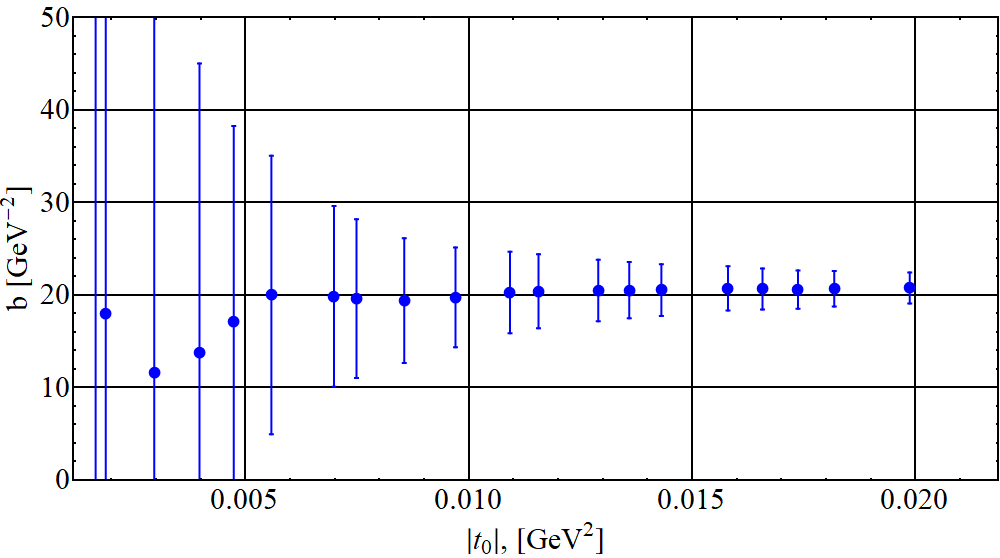}$$
\caption{The values of the parameter $b$  and its error, depending
on the cut-off value $|t_{0}|$ of the experimental data  with full
errors. However, in all these results, $\chi^{2}/\mbox{NoF} \ll 1$, and
these results should be regarded as evaluative. }
\label{fig-6b}       
\end{figure}

Similar statements hold for the other fit parameters: $(d\sigma/dt)_{t=0}$
and $b$. Their behaviour is presented in figs.~\ref{fig-5a}, \ref{fig-5b},
\ref{fig-6a}, \ref{fig-6b}.

\begin{figure}
$$\includegraphics[width=140mm]{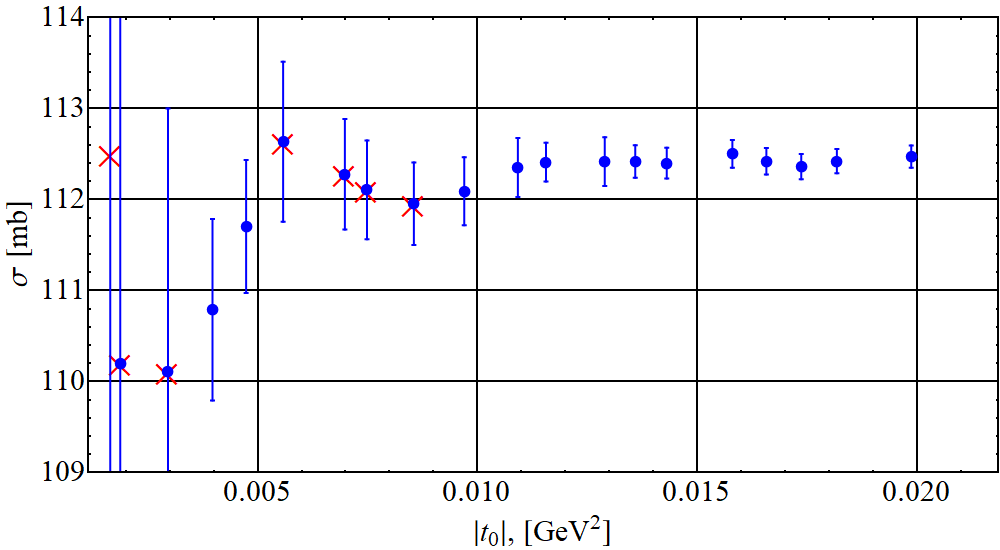}$$
\caption{The values of $\sigma_{\mbox{tot}} $ and its error depending
on the cut-off value $|t_{0}|$ of the experimental data  taking
into account only statistical errors. Crossed out are the results for
which $\chi^{2}/\mbox{NoF}>1 $.}
\label{fig-7a}       
$$\includegraphics[width=140mm]{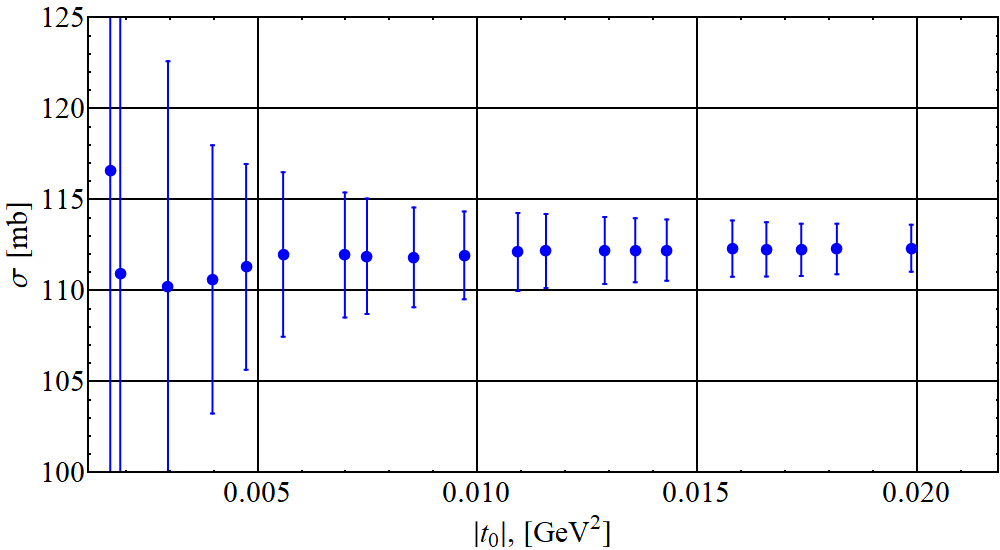}$$
\caption{The values of $\sigma_{\mbox{tot}} $ and its error, depending
on the cut-off value $|t_{0}|$ of the experimental data  with
full errors. However, in all these results, $\chi^{2}/\mbox{NoF} \ll 1$,
and these results should be regarded as evaluative.}
\label{fig-7b}       
\end{figure}

Having the values of these parameters, it is possible to estimate the total
cross-section by the formula:\vspace{-3.0mm}
\begin{equation}
\sigma_{\mbox{tot}} =
4\hbar c\sqrt{\pi (d\sigma/dt)_{t=0}}\vspace{-3.0mm}
\label{eq3}
\end{equation}
In the figures figs. \ref{fig-7a} and \ref{fig-7b}
shows the $\sigma_{\mbox{tot}}$ corresponding to these parameters.

The behaviour of the total cross section and its errors depending on the cut-off value $ t_0 $ leads to the following values of the total cross
sections:\vspace{-3.0mm}
$$
\sigma_{\mbox{tot}} =110.0\pm 8.0 ~[\mbox{mb}] ~~~
\mbox{(including complete errors),~~~~~~~~~~~~~~~~}\vspace{-5.0mm}
$$
$$
\sigma_{\mbox{tot}} =110.8\pm 1.0 ~[\mbox{mb}] ~~~
\mbox{(with account of statistical errors only).}\vspace{-7.0mm}
$$

\section*{Evaluation of parameters when accounting for correlations of experimental data.\vspace{-3.0mm}} 

When fitting using error sources, it becomes possible to use all five parameters in formula (5).
Fig. 8, 9, 10 show the results of fitting with three, two, and one parameter in the exponent, respectively. Blue color indicates the results with all error sources, and the red means all error sources except normalization (except for the first column in the error source).
In these two cases, the results no longer differ so dramatically as in the previous paragraph, although of course one needs to trust the results when taking into account all the sources of errors.
Plots showing $\chi^{2}/$NoF are also provided in order to understand which results should be discarded.

The final results of the description with error sources are summarized in tables  to be shown below.\vspace{-7.0mm}

\section*{Summary of the parameters from describing the data with the "TOTEM model" for $ T_{N}(s,t) $ and modified account of CNI\vspace{-3.0mm}}

Although we present the values of the fitting parameters  without taking into account
normalization, we do not at all consider this method of action to be completely
correct because if there is a source of errors, then it must be taken into
account (why then do the experimenters bring them?).
From the tables and graphs below, it also follows that the use of more than
one term in the exponent (5)
$( b_{1}|t|+ b_{2}|t|^{2} + b_{3}|t|^{3})$ is redundant and  one term
$b_{1}|t| \equiv b |t|$ is fairly enough to get the same
value of $\chi^{2}/\mbox{NoF}$. In addition, there is a strong correlation
among all the three.

The smallest value for  $\rho$ is obtained for option 1 (see above: all three
parameters in the exponent): 
\begin{center}
$\rho(3b_{i}) = 0.104 \pm 0.022 $ . 

\end{center}

However, we believe that due to the excessive number of parameters
$ b_{i}$, there is no good reason for accepting such a result.

The totality of all fit results indicates that there is no statistically
significant reasons to fall below $ t_ {0} = $ 0.015 GeV$^{2}$, because in
these cases either $ \chi^2 /\mbox{NoF}$ is too close to zero (which is statically
unacceptable), or $\chi^2 /\mbox{NoF} > 1$, which gives a statistical certainty
of less than 50{\%}. In Tables 1 and 2 we give a general summary of the
parameters and their errors (as well as their correlation matrices) for cases
of accounting the source of normalization errors (the largest contribution
to the error in the experimental data table in \cite {Ant}) and without it,
respectively \footnote {In all these
cases, the expected value of $\sigma_{tot}$ is given  according to the
formula (3).}. All calculations are done for three options of the number
of parameters $b_{i}$ in the exponent:
\vspace {-3.0mm}
\begin{enumerate}
\item a single parameter  $b_{1}$;\vspace{-3.0mm}
\item two parameters  $b_{1}$ and $b_{2}$;\vspace{-3.0mm}
\item three parameters $b_{1}$, $b_{2}$ and $b_{3}$.\vspace{-3.0mm}
\end{enumerate}

The results obtained without taking into account the source of normalization errors, despite that they give in some cases the expected values of
$\rho <$ 0.1, are hardly worth considering (Table 2).
Since experimenters give this source of error, there is no reason
not to use them when numerically processing the experimental results.

Thus, we are inclined to believe that parameter values should be taken
from Table 2.

Once again, we emphasize, based on this table, that there is no need to
use two (moreover, three) parameters in the exponent, because
even  without them we get a statistically significant $\chi^2/ $ NoF. Besides, this would also result in  a sharp increase in the error values of these parameters.

{\it Thus, based on the considerations presented above, we believe that
the correct result for the parameter $\rho$  obtained using all sources
of errors, the modified formula for CNI and with only the linear term  in
the exponent} ( $b|t|/2$) {\it of the} "TOTEM" {\it model for $T_{N}$ reads}:
\begin{equation}
\boldmath{
\rho(13~\mbox{TeV}) = 0.123 \pm 0.010,}  \label{eq4}
\end{equation}
\textit{what corresponds to the following value of the total cross section:}
\begin{equation}
\boldmath{
\sigma_{\mbox{tot}}(13~\mbox{TeV}) = 111.4 \pm 1.8 ~[\mbox{mb}].
}\label{eq5}
\end{equation}
\begin{figure}
\noindent $\includegraphics[width=175mm]{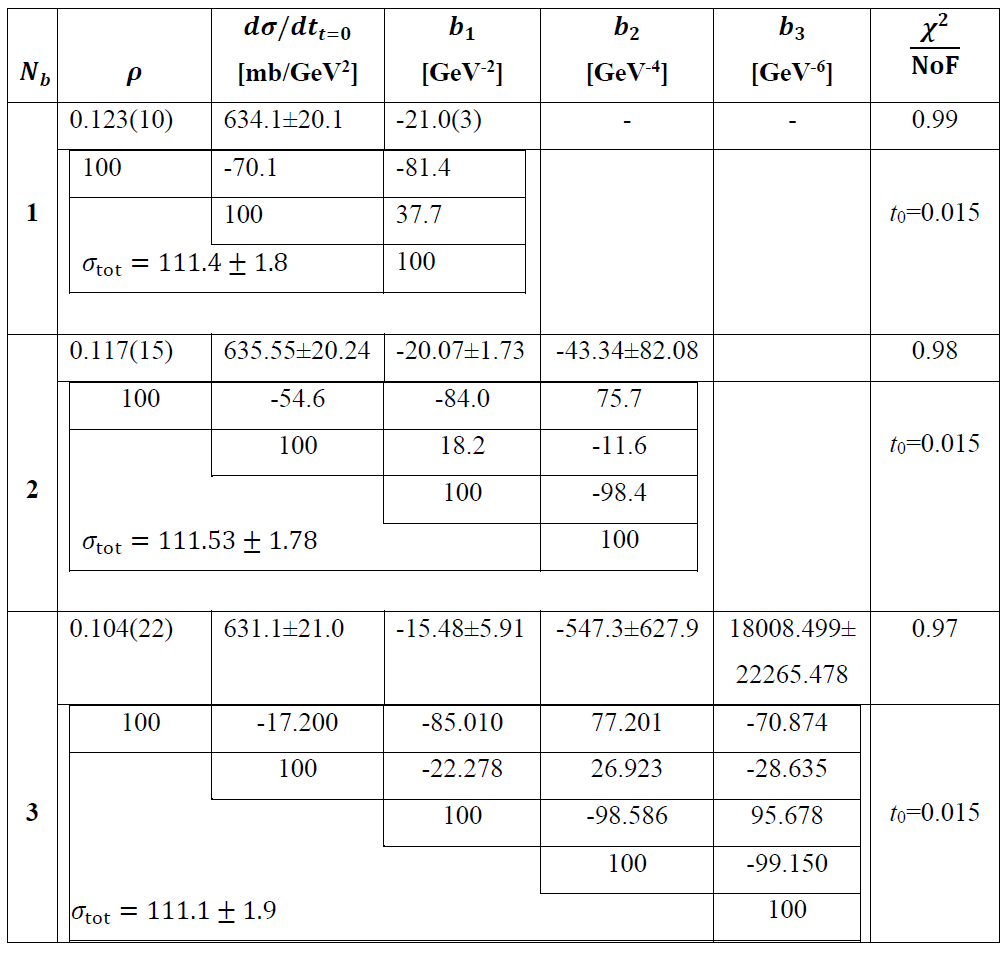}$
\caption{Table 1. Parameter values with all error sources}
\label{Table1}       
\end{figure}
\begin{figure}
\noindent $\includegraphics[width=175mm]{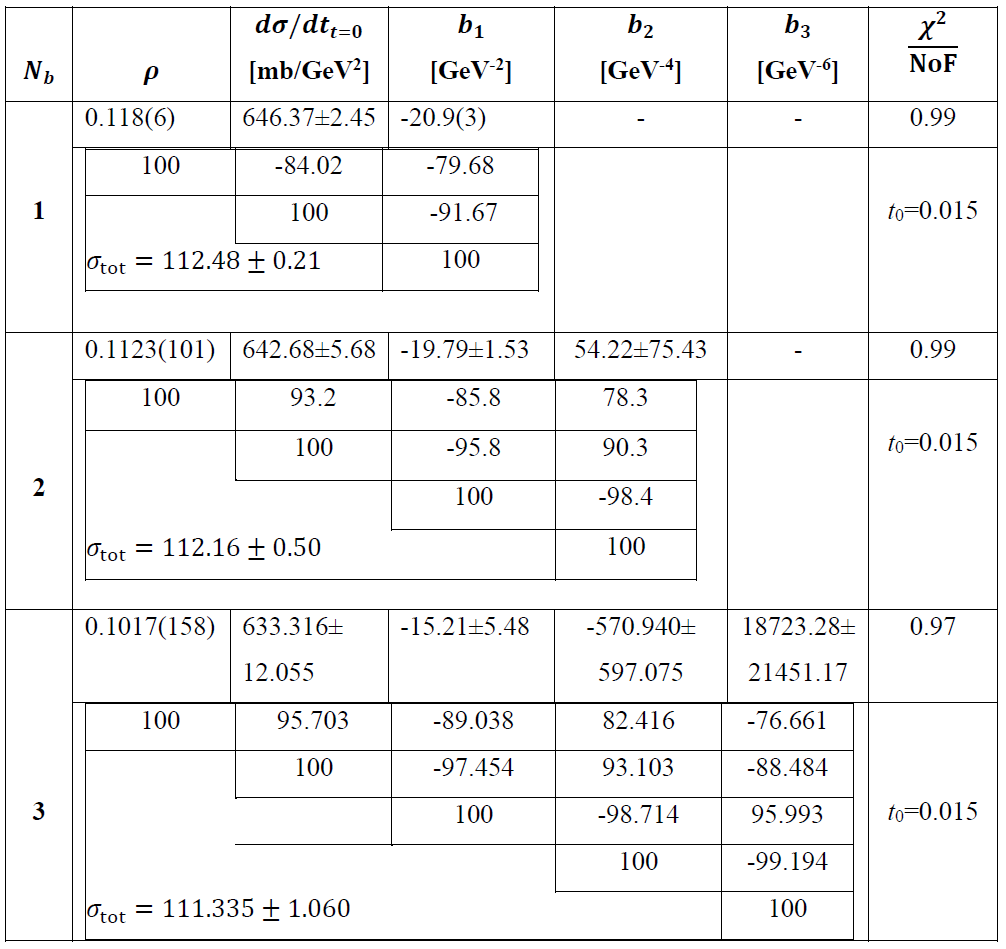}$
\caption{Table 2. Parameter values with all error sources except normalization.}
\label{Table2}       
\end{figure}
\newpage

\section*{Whether the COMPETE parametrization is ruled out?\vspace{-3.0mm}}
Here we would like to comment in passing the following categorical
judgement made in \cite{Ant}: 
"...{\it none of the} COMPETE {\it models is compatible with the ensemble
of} TOTEM's $\sigma_{\mbox{tot}}$ {\it and} $\rho$ {\it measurements.}"
In simple words this means that the use of the COMPETE parametrizations as
given in \cite{COMP}  cannot lead to simultaneous description of
$\sigma_{\mbox{tot}}$ and $\rho$ which is consistent with the values of these
quantities  given in \cite{Ant}.

This issue has been addressed in \cite{Cud}  with the conclusion that the
actual predictions of the COMPETE modelling set for $\sqrt{s}=$
13~TeV is the whole band:\vspace{-7.0mm}

$$86~\mbox{mb} \leq \sigma_{\mbox{tot}} \leq 117 ~\mbox{mb}\vspace{-4.0mm}$$
and\vspace{-4.0mm}
$$0.058\leq \rho \leq 0.145.\vspace{-2.0mm}$$

\begin{figure}
$$\includegraphics[width=175mm]{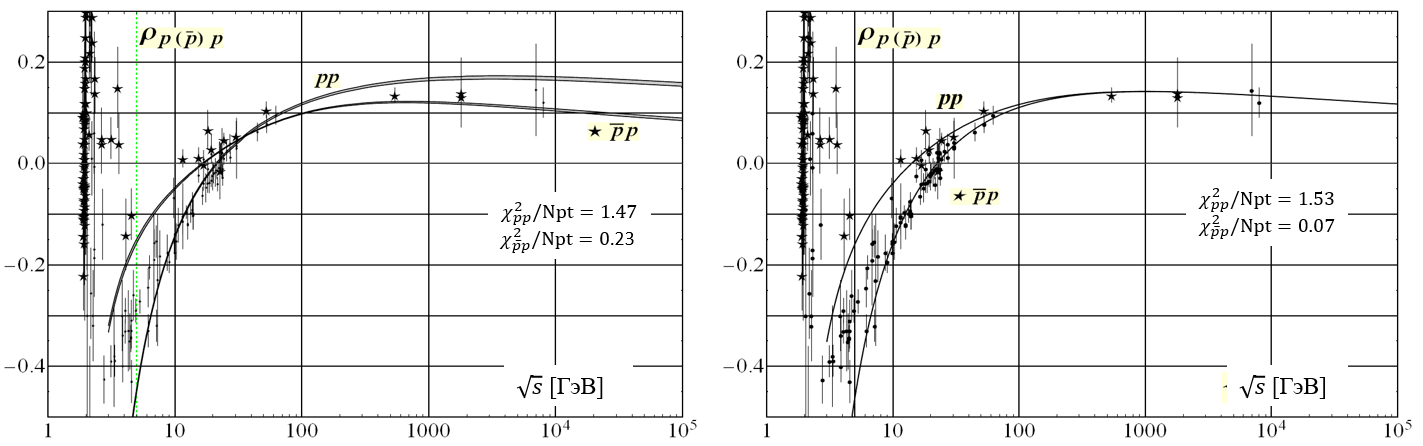}\vspace{-5.0mm}$$
\caption{Evolution of the  $\rho$ parameter with the maximal Odderon (left)
and without it (right).}
\label{With_Odderon_and_without_Odd}       
\end{figure}

The only thing that could be concluded (if we would take for granted the
values of $ \sigma_{\mbox{tot}} $ and $\rho $ given in \cite{Ant}) is that
the fit which was considered earlier as the best one, should be accommodated
to new data. This opportunity is in no way excluded by the general framework
of COMPETE.
In terms of discussing of the presence or absence of the Odderon, mention
should be made of the behaviour of the curves of the $\rho$-parameter
in these two cases. We carried out a study of the description of all experimental
data of differential cross sections
in the range from 7 GeV to 13 TeV together with experimental data on
$ \sigma_{\mbox {tot}} $ and $ \rho $ parameter between 5 GeV and 13 TeV
\cite{Av}. Fig.~\ref{With_Odderon_and_without_Odd} shows the behaviour of
these curves. It is seen that in the case of the presence of the Odderon
the curve of the $ \rho $ parameter passes even higher than $ 0.123 $  at
$\sqrt {s}= 13$ TeV. In addition, in the case of vanishing maximum Odderon
contribution, the curves for $ pp $ and $ {\bar p} p $ converge at high energies
as is observed in the COMPETE model. This fundamentally contradicts the behaviour
of these curves in the case of the presence of the Odderon.
\section*{On slowing down of the total cross-section growth.\vspace{-3.0mm}}
It was announced in  \cite{Ant}    that \textit{...the  $ \rho $ value determined
by TOTEM would represent a first evidence of a slowing down of the total
cross-section growth at higher energies.}
 It would be instructive
to try to understand whether some specific value of $ \rho $ would imply
a decrease in the rate of growth of $ \sigma_{\mbox{tot}} $. For this purpose
let us use the derivative forward dispersion relations.
Differential dispersion relations read \cite{Su} (for more rigorous treatment
see Ref.\cite{Kol})\vspace{-0.0mm} 
\begin{equation}
\frac{\mbox{Re}[T(s,0)]}{s} = \frac{\pi}{2}\frac{d}{dy}
\left( \frac{\mbox{Im}[T(s,0)]}{s} \right) ,
~~~y=\ln s. \label{eq9}\vspace{-0.0mm}
\end{equation}
The "slowing down of the total cross-section growth" announced in \cite{Ant})
means that at the LHC  energy $\mathcal{O}$$(\mbox{13 TeV})$ when
$\rho = 0.1$ the acceleration $d^{2}\sigma_{\mbox{tot}}/dy^{2}$ vanishes
and changes from positive to the negative one. 
It is not a hard task to derive from Eq.(9) that equation\vspace{-4.0mm}
$$d^{2}\sigma_{\mbox{tot}}/dy^{2} = 0~~
\mbox{is equivalent to equation}~~
d(\rho\cdot\sigma_{\mbox{tot}})/dy = 0.\vspace{-3.0mm}
$$
If to take the values of the product $(\rho\cdot\sigma_{\mbox{tot}})$ 
up to 13 TeV, as presented in the TOTEM publications,  it is  seen that the value of
$\rho\cdot\sigma_{\mbox{tot}}$ really reaches its maximum 
in the vicinity of 7 TeV .
This would mean that already at $\sqrt{s} \simeq$ 7 TeV the total cross section started
to decelerate (although still increasing).
 
It is clear that this conclusion is tightly related with the low value of $\rho$ announced by the TOTEM Collaboration in \cite{Ant}.

On the other hand, according to the COMPETE parametrization \cite{COMP}
the value  $ \rho = 0.1 $ will be reached only at
$\sqrt{s}=$ $\mathcal{O}$$(10^{6})$ TeV while the value of
$(\rho\cdot\sigma_{\mbox{tot}})$ steadily grows and at present energies does
not reveal any sign of saturation.

The same conclusion follows from our estimations
(\ref{eq4}) and (\ref{eq5}).\vspace{-9.0mm}

\section*{Summary and conclusions.\vspace{-4.0mm}}
We have reanalyzed the TOTEM data at 13 TeV \cite{Ant} on
the basis of a new treatment of CNI but with the same nuclear amplitude.
Our results significantly differ from those presented in \cite{Ant}:\vspace{-3.5mm}
\begin{enumerate}
\item TOTEM \cite{Ant}: {\boldmath $\rho =0.09\pm 0.01~ (0.10\pm 0.01);~
\sigma_{\mbox{tot}}= 110.5\pm 2.4 ~\mbox{mb}.$};\vspace{-4.0mm}
\item This paper:~ {\boldmath ${\rho = 0.123\pm 0.010; ~~~~~~~~~~~~~~~~~~
\sigma_{\mbox{tot}}=
111.4\pm 1.8 ~\mbox{mb}}.\vspace{-3.0mm}$}
\end{enumerate}
 We also presented arguments against a supposed total failure of the COMPETE framework claimed in \cite{Ant} and have shown that the conclusion about slowing down of the total cross-section growth critically depends on the values of the product $(\rho\cdot\sigma_{\mbox{tot}})$. According to our results such a slowing down does not take place at present energies.

\newpage

\section*{Appendix.}

\section*{TOTEM data and the Bethe formula for CNI.}\vspace{-2.0mm}

Classical results on the $\rho$-parameter and the total cross section
$\sigma_{\mbox{tot}}$ at the ISR  were presented in [8]. However, in this
work, to extract the $ \rho $-parameter, a different (from that of used in
the TOTEM publication) formula for account of CNI was used (the well known
Bethe parametrization) which gives:\vspace{-1.0mm}
\begin{equation}
\boldmath{
\frac{d\sigma}{dt}= \frac{4\pi g \alpha^{2}F^{4}(t)}{t^{2}}+
\frac{\sigma_{\mbox{tot}}\alpha[\rho +\alpha\phi(t)F^{2}(t)]e^{bt/2}}{t}+
\frac{(1+\rho^{2})\sigma_{\mbox{tot}}^{2} e^{bt}}{16\pi g}} \label{eq7}\vspace{-2.0mm}
\end{equation}
where\vspace{-2.0mm}
$$\phi(t) = \ln \left( \frac{0.08}{-t}\right) - 0.5772...~.\vspace{-2.0mm}$$
The fitting parameters are  $\sigma_{\mbox{tot}}$, $\rho$ and $b$.

We applied this formula to the TOTEM data at 13 TeV, slightly modifying its
mathematical form and taking the expression for  $\sigma_{\mbox{tot}}$
used in the "TOTEM model" in eq. (\ref{eq3}).

So, we have got again three parameters: $(d\sigma/dt)_{t=0}$,
$\rho$ and $b$. 

\begin{figure}
$$\includegraphics[width=150mm]{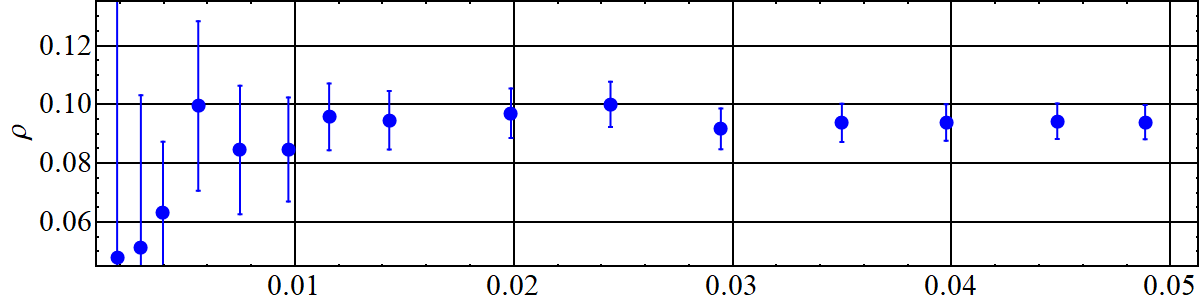}$$\vspace{-8.0mm}
$$\includegraphics[width=150mm]{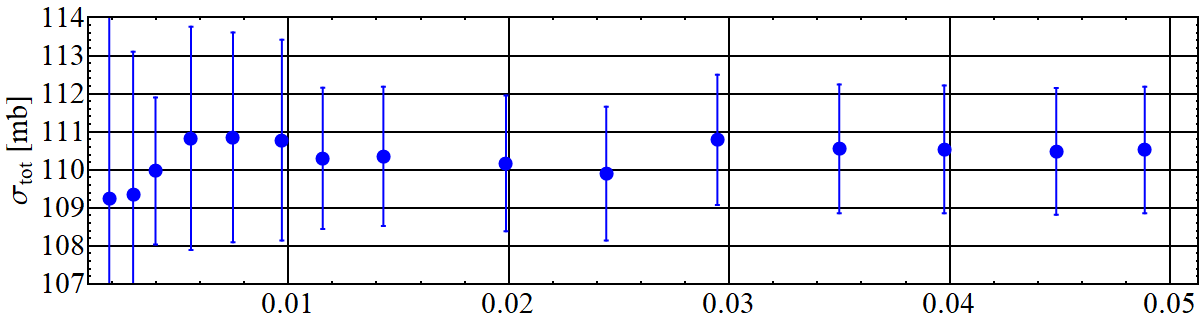}$$\vspace{-8.0mm}
$$\includegraphics[width=150mm]{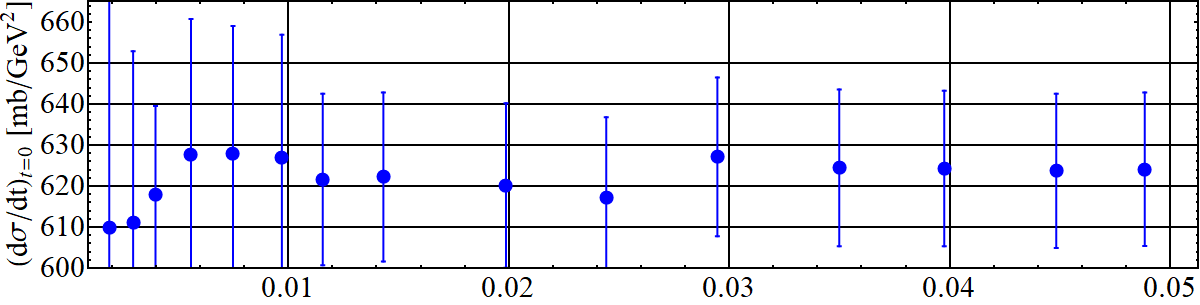}$$\vspace{-8.0mm}
$$\includegraphics[width=150mm]{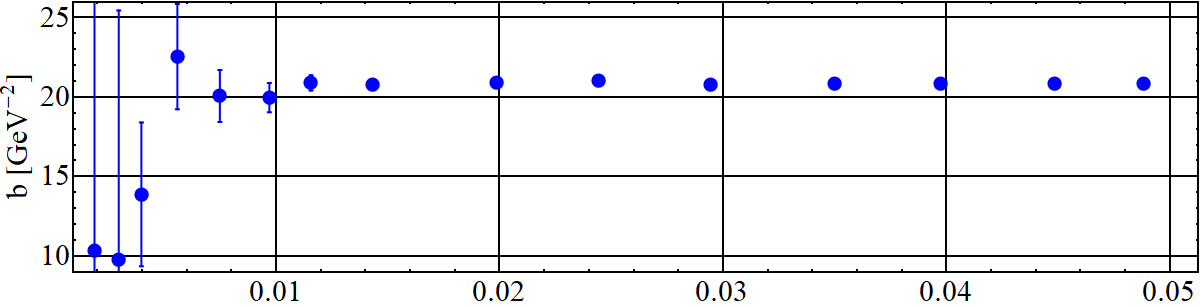}$$\vspace{-8.0mm}
$$\includegraphics[width=150mm]{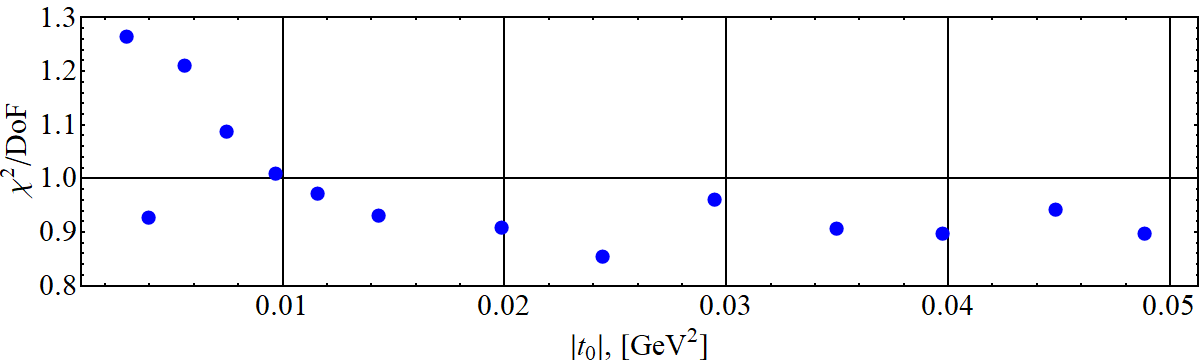}$$\vspace{-13.0mm}
\caption{ Parameters at 
$\sqrt{s}=13$ TeV  (\ref{eq7}) as function of the cut-off $t_{0}$.}
\label{fig-ISR}       
\end{figure}

Fig. \ref{fig-ISR} shows the results of processing the experimental data
according to the procedure that we used earlier. It can be seen that
the value $\chi^{2}/$NoF stably lies in a small neighbourhood of unity
(and remains less than this value) until the cut-off of the experimental
points\footnote{ As was obtained earlier,
the introduction of additional parameters $ b_{2,3} $ in the exponent is
redundant.} with $|t|> t_0 \cong 0.012 GeV^{2}$.

Based on this, we come to  the following parameter values:\vspace{-4.0mm}
$$
\boldmath{
\begin{array}{rcl}
\rho&=&0.0958 \pm 0.0113;\\[-0.0mm]
\sigma_{\mbox{tot}}&=&110.3\pm 1.9~[\mbox{mb}];\\[-0.0mm]
(d\sigma/dt)_{t=0}&=&632.6\pm 20.9~[\mbox{mb/GeV}^{2}].\\[-0.0mm]
\end{array}
} \vspace{-2.0mm}
$$
These results are very close to the values declared by the TOTEM collaboration
\cite{Ant}. Note that we also tested the fitting procedures on experimental
data obtained at ISR energies and at lower energies (the U-70 Serpukhov
data [9]). The results are very close to the values presented in
these works.\vspace{-1.0mm}

Thus, we see that the values of the $\rho $ parameter extracted from
experimental data are significantly model-dependent.
This alone negates the conclusion of \cite{Ant} that the COMPETE model
does not correctly describe the behaviour of the $\rho $ parameter.

\newpage
\newpage


\end{document}